\documentclass[twocolumn]{afmc_art}
\usepackage{graphicx}
\begin{document}

%
%
%
%
%
%
%
%
\title{Pre-unstable set of multiple transient three-dimensional perturbation waves and the associated turbulent state in a shear flow}
%
%
%
\author{S. Scarsoglio$^1$ and D. Tordella$^2$}
\affiliation{$^1$Department of Hydraulics\\
             Politecnico di Torino, Torino 10129, Italy\\[5pt]
             $^2$Department of Aeronautics and Space Engineering\\
             Politecnico di Torino, Torino 10129, Italy}

\maketitle
%
%
\section{Abstract}


In order to understand whether, and to what extent, spectral representation can effectively highlight the nonlinear interaction among different scales, it is necessary to consider the state that precedes the onset of instabilities and turbulence in flows. In this condition, a system is still stable, but is however subject to a swarming of arbitrary three-dimensional small perturbations. These can arrive any instant, and then undergo a transient evolution which is ruled out by the initial-value problem associated to the Navier-Stokes linearized formulation. The set of three-dimensional small perturbations constitutes a system of multiple spatial and temporal scales which are subject to all the processes included in the perturbative Navier-Stokes equations: linearized convective transport, linearized vortical stretching and tilting, and the molecular diffusion. Leaving aside nonlinear interaction among the different scales, these features are tantamount to the features of the turbulent state.

We determine the exponent of the inertial range of arbitrary longitudinal and transversal perturbations acting on a typical shear flow, i.e. the bluff-body wake. Then, we compare the present results with the exponent of the corresponding developed turbulent state (notoriously equal to $-5/3$). For longitudinal perturbations -- i.e. perturbations in the plane of the basic flow which is two-dimensional -- we observe a decay rate of $-3$ in the inertial range, typically met in two-dimensional turbulence. For purely three-dimensional perturbations, instead, the energy decreases with a factor of $-5/3$. If we consider a combination of longitudinal and transversal perturbative waves, the energy spectrum seems to have a decay of $-3$ for larger wavenumbers ($k\in[50, 100]$), while for smaller wavenumbers ($k\in[3,50]$) the decay is of the order $-5/3$. We can conclude that the value of the exponent of the inertial range has a much higher level of universality, which is not necessarily associated to the nonlinear interaction.

\section{Introduction}

One  very popular notion in the phenomenology of turbulence (in the sense of Kolmogorov 1941) is that a power-law scaling with an exponent close to $-5/3$ is observed for the energy spectrum over a very substantial range of a few decades of wavenumber, this range being called the inertial range. For extensive collections of laboratory and numerical experimental results, see for instance \cite{F95,SA97}. In fact, it is a common criterion for the successful production of a fully developed turbulent field, either in the laboratory or in numerical simulations, to verify that the power spectrum has such a scaling in the inertial range.

The set of arbitrary three-dimensional small perturbations constitutes a system of multiple spatial and temporal scales which are subject to all the processes included in the  Navier-Stokes equations, leaving aside the nonlinear interaction among the different scales.
If it were possible to observe such a system in a temporal window and obtain the instantaneous power spectrum, it would be possible, among others, to determine the exponent of the inertial range of the arbitrary perturbation, and to compare it with the exponent of the corresponding developed turbulent state. Two possible situations can therefore appear: (a) the exponent difference is large and, as such, it is a quantitative measure of the nonlinear interaction in spectral terms; (b) the difference is small. This would be even more interesting, because it would indicate a higher level of universality on the value of the exponent of the inertial range,  not necessarily associated to the nonlinear interaction.

We propose building temporal observation window for the transient evolution of a large number (order of $10^2 \sim 10^3$) of  arbitrary small 3D (oblique) perturbations acting on a typical shear flow. In particular, we can take advantage of a  recently -- numerically obtained -- set of solutions yielded by the initial-value problem applied to 3D perturbations of a plane bluff-body wake \cite{STC09,STC10}. These solutions have revealed the existence of  many different kinds of transient behaviour, not all of which is trivial. If these transients, obtained in association with arbitrary initial conditions, are injected in a statistical way into the temporal observation window, we can obtain a close representation of the perturbation state that precedes the onset of instability-turbulence. In this preliminary work, we consider an ensemble of transients of asymptotically stable waves -- this is because we want to avoid any temporal divergence -- and build the energy spectrum by freezing each wave at the moment it reaches a constant value of the temporal growth rate. Here, the growth is always negative since all waves are asymptotically damped in time. When this happens, the wave may be considered out of its transient and in the asymptotic condition. When building the power spectrum, the energy of each wave is normalized over its initial energy. In this way, we are considering the spectral dynamics of a sort of white noise perturbation made up of asymptotically stable waves.

\section{Formulation}

\begin{figure}
\begin{minipage}[]{\columnwidth}
   \includegraphics[width=\columnwidth]{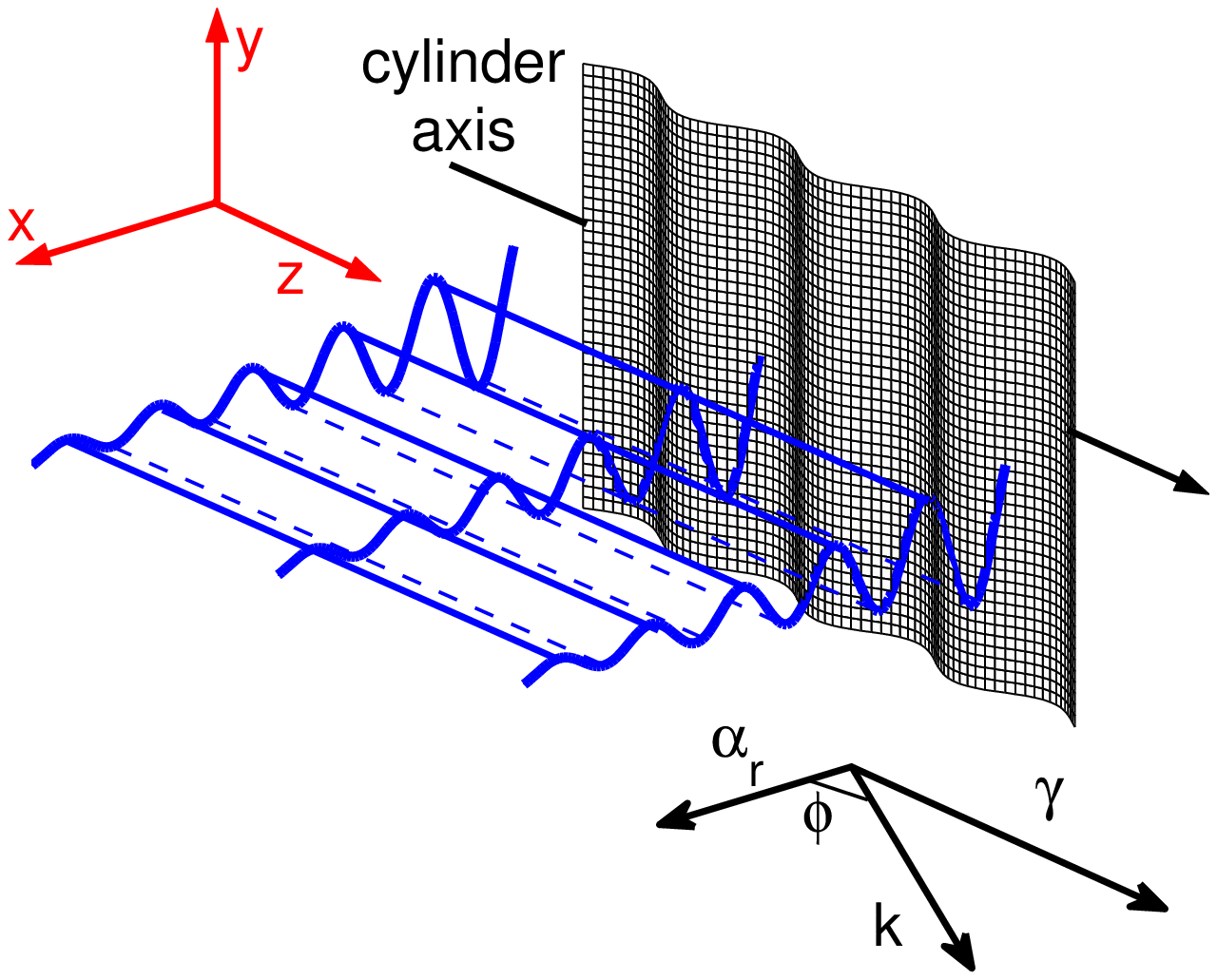}
    \label{G_k}
\vspace{-2.4cm}
\end{minipage}
\begin{minipage}[]{0.5\columnwidth}
   \includegraphics[width=\columnwidth]{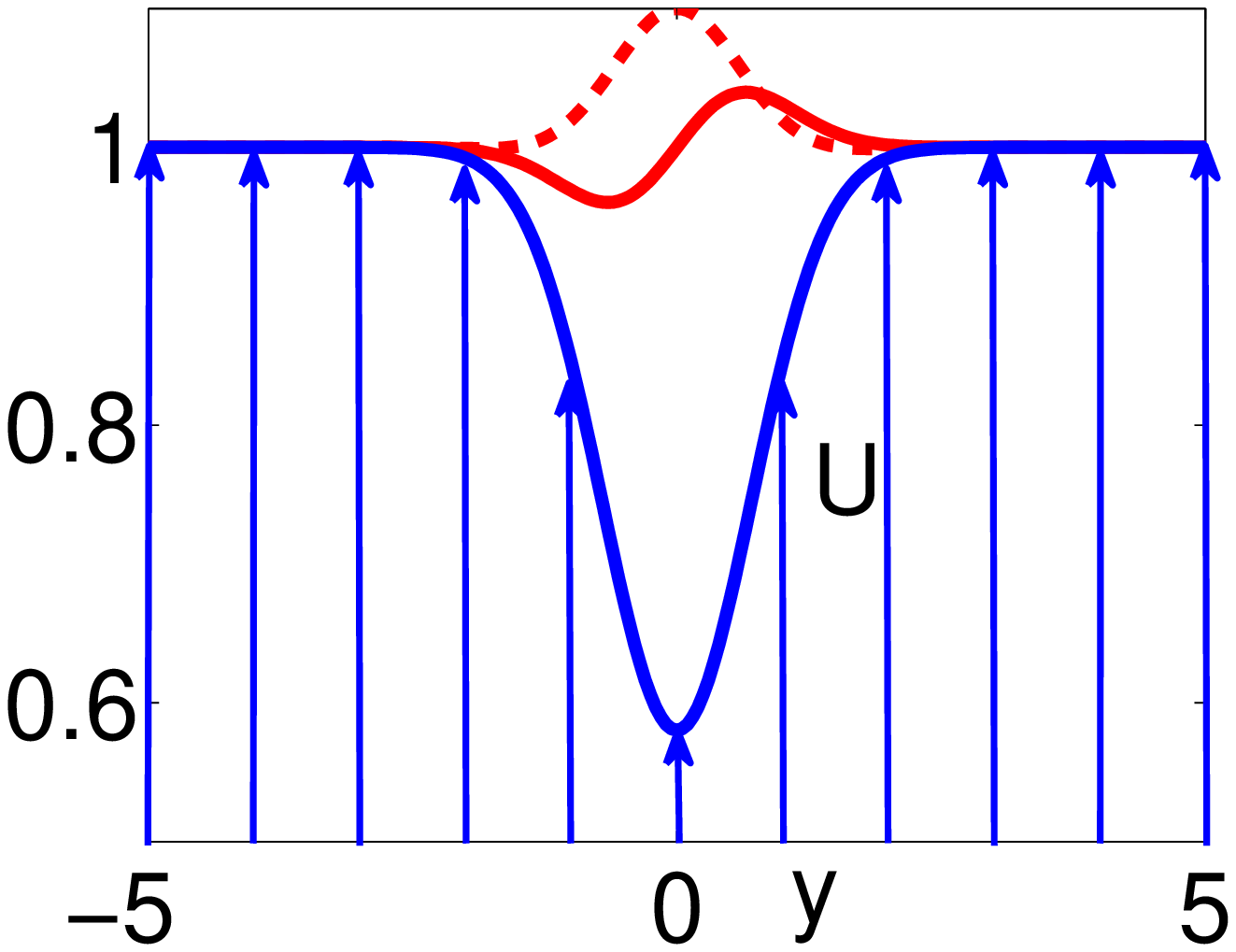}
    \label{G_k}
\end{minipage}
  \caption{Perturbation geometry scheme, mean flow $U$ in the wake at $Re=40$ and $x_0=10$ (blue curve), symmetric and asymmetric initial conditions in terms of $\hat{v}(t=0,y)$ (red curves).}\label{perturbation_scheme}
\end{figure}

\noindent The energy spectrum behaviour is studied using the initial-value problem
formulation. The base flow is approximated at a fixed longitudinal station, $x_0=10$, through an
analytical expansion solution \cite{TB03} of the Navier-Stokes
equations (see Fig. \ref{perturbation_scheme}). The Reynolds number is set to a value of $40$, in order to consider stable evolutive configurations. The viscous perturbative equations are written in terms
of the vorticity and the transversal velocity \cite{CD90} and then
transformed through a Laplace-Fourier decomposition \cite{STC09,STC10}
in the plane ($x, z$) which is normal to the base flow plane ($x,
y$),

\begin{eqnarray} \label{IVP2_fou1}
\frac{\partial^2 \hat{v}}{\partial y^2} &-& (k^2 - \alpha_i ^2 + 2
i \alpha_r \alpha_i) \hat{v}= \hat{\Gamma}, \\
\nonumber \frac{\partial \hat{\Gamma}}{\partial t} &=& (i \alpha_r -
\alpha_i) (\frac{d^2 U}{dy^2} \hat{v} - U \hat{\Gamma}) +\\
&+& \frac{1}{Re} [\frac{\partial^2 \hat{\Gamma}}{\partial y^2} - (k^2
- \alpha_i ^2 + 2 i \alpha_r \alpha_i)
\hat{\Gamma}],\label{IVP2_fou2} \\
\nonumber \frac{\partial \hat{\omega}_{y}}{\partial t} &=& - (i \alpha_r - \alpha_i) U \hat{\omega}_{y}
  - i \gamma \frac{dU}{dy} \hat{v}+\\
 &+& \frac{1}{Re} [\frac{\partial^2 \hat{\omega}_y}{\partial
y^2} - (k^2 - \alpha_i ^2 + 2 i \alpha_r \alpha_i)
\hat{\omega}_y].\label{IVP2_fou3}
\end{eqnarray}

\noindent  The transversal velocity and vorticity components are indicated as $\hat{v}$ and $\hat{\omega}_{y}$ respectively, while $\hat{\Gamma}$ is defined through the kinematic relation  $\widetilde{\Gamma} = \partial_x \widetilde{\omega}_z - \partial_z \widetilde{\omega}_x
$ that in the physical plane links together the perturbation vorticity
components in the $x$ and $z$ directions ($\widetilde{\omega}_x$
and $\widetilde{\omega}_z$) and the perturbed
velocity field. Equations (\ref{IVP2_fou2}) and
(\ref{IVP2_fou3}) are the Orr-Sommerfeld and Squire equations
respectively, from the classical linear stability analysis for
three-dimensional disturbances in the phase space. We define $k$ as the
polar wavenumber, $\alpha_r = k cos(\phi)$ as the wavenumber in $x$ direction, $\gamma = k sin(\phi)$ as the
wavenumber in $z$ direction, $\phi$ as the angle of
obliquity with respect to the physical plane, and $\alpha_i$ as the
spatial damping rate in $x$ direction (see the perturbation scheme in Fig. \ref{perturbation_scheme}). The measure of the perturbation growth can
be defined through the disturbance kinetic energy density in the plane $(\alpha, \gamma)$:

\begin{eqnarray}
\label{kinetic_energy} &&e(t; \alpha, \gamma) =
\int_{-y_d}^{+y_d} (|\hat{u}|^2
+ |\hat{v}|^2 + |\hat{w}|^2) dy =\\
\nonumber &=& \frac{1}{|\alpha^2 +
\gamma^2|}\int_{-y_d}^{+y_d} \left(\left|\frac{\partial \hat{v}}{\partial
y}\right|^2 + |\alpha^2 + \gamma^2| |\hat{v}|^2 + |\hat{\omega}_y |^2\right)
dy,
\end{eqnarray}

\noindent where $\hat{u}$ and $\hat{w}$ are the streamwise and spanwise components of the perturbation velocity, respectively, while $2y_d$ is the extension of the spatial numerical
domain. The amplification factor $G(t)$ can be
introduced in terms of the normalized energy density

\begin{figure}[h!]
\centering
\begin{minipage}[]{0.8\columnwidth}
   \includegraphics[width=\columnwidth]{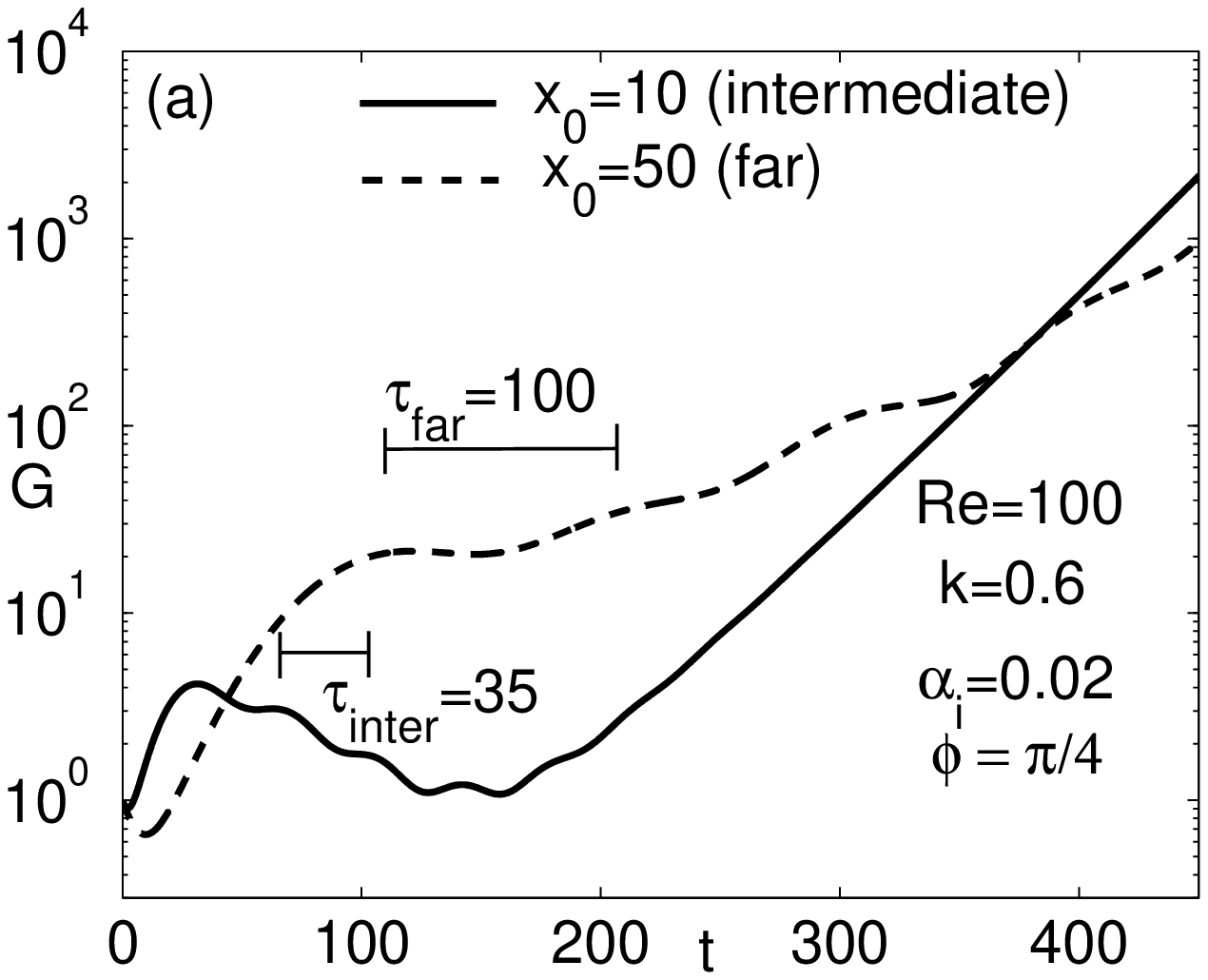}
    \label{G_k}
\end{minipage}
\begin{minipage}[]{0.8\columnwidth}
   \includegraphics[width=\columnwidth]{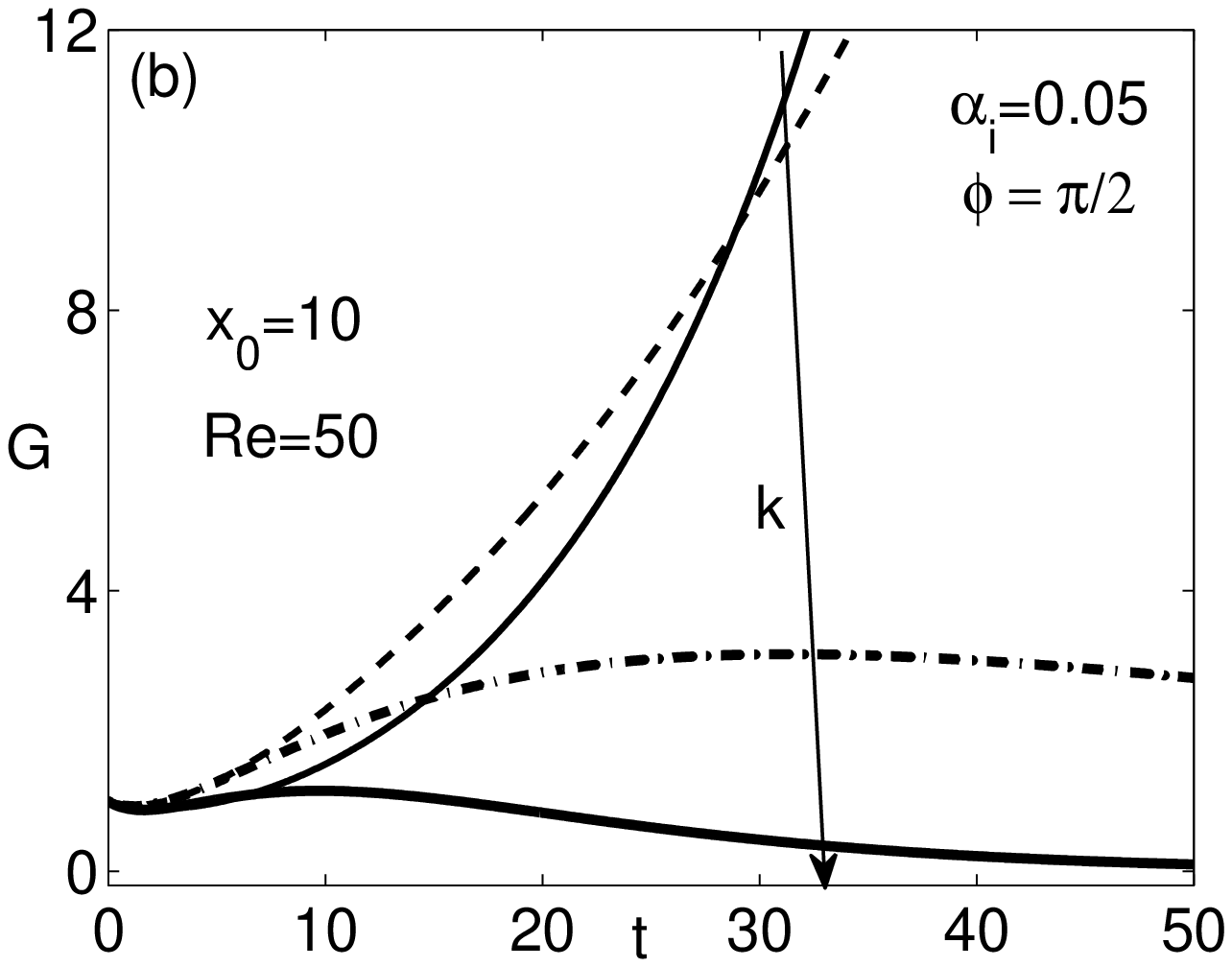}
    \label{r_k}
\end{minipage}
\begin{minipage}[]{0.8\columnwidth}
   \includegraphics[width=\columnwidth]{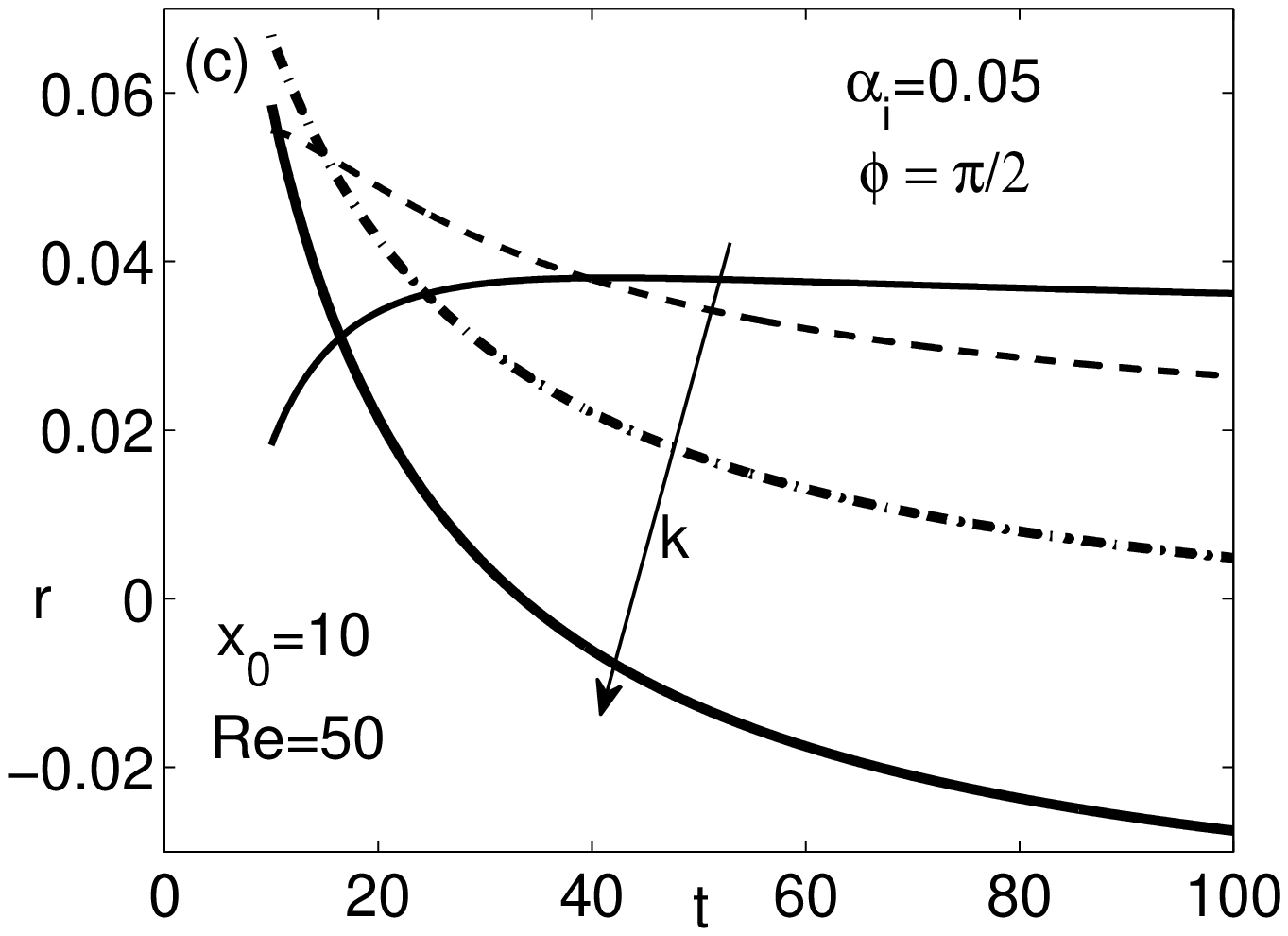}
    \label{r_k}
\end{minipage}
\caption{(a) The
amplification factor $G$, asymmetric
initial condition, intermediate ($x_0=10$, solid curves) and far field ($x_0=50$,
dashed curves) wake configurations. The periods $\tau_{inter},
\tau_{far}$ are the periods of the modulation visible on $G$, in
the intermediate and far field, respectively. (b) The amplification
factor $G$ and (c) the temporal growth rate $r$ as function of
time, symmetric initial condition, $k = 0.5, 1, 1.5, 2$.}
\label{examples_transient}
\end{figure}

\begin{equation}
G(t; \alpha, \gamma) = \frac{e(t; \alpha, \gamma)}{e(t=0; \alpha,
\gamma)}.
\end{equation}

\noindent The temporal growth rate on the kinetic energy $r$

\begin{equation}
r(t; \alpha, \gamma) = \frac{log|e(t; \alpha, \gamma)|}{2t},
\;\;\; t>0 \label{IVP2_tgr}
\end{equation}

\noindent is introduced in order to evaluate both the early
transient as well as the asymptotic behaviour of the
perturbations.

Examples, in terms of $G$ and $r$, of significant transient behaviour
and asymptotic fate of the three-dimensional perturbations \cite{STC09} are shown in Fig. \ref{examples_transient}. In part (a) the amplification factor $G$ of an asymmetric wave is shown for two typical intermediate ($x_0=10$) and far ($x_0=50$) wake configurations. For $x_0=10$ a
local  maximum, followed by a minimum, is visible in the energy
density, then the perturbation is slowly amplifying and the
transient can be considered extinguished only after hundreds of
time scales. For $x_0=50$ these features are less marked, but still present. For both configurations, the function $G$ shows a modulation in time in the first part of the transient ($\tau_{far}=100$ and $\tau_{inter}=35$), which is always observed in the case of asymmetric
longitudinal or oblique instability waves. In parts (b) and (c) of Fig. \ref{examples_transient}, the amplification factor $G$ and the temporal growth rate $r$ of purely three-dimensional symmetric perturbations are reported, respectively. Such waves may become damped by
increasing their wavenumber ($k=\gamma$). Before the asymptotic stable states are reached,
these configurations yield maxima of the energy density (e.g. when
$k=1.5, G \sim 3$ at $t\sim 30$) in the transients. This trend is
also typical of oblique and longitudinal waves, and it can be
considered a general feature in the context of the stability of
near parallel shear flows.

\section{Results}

\begin{figure}
\begin{minipage}[]{\columnwidth}
   \includegraphics[width=\columnwidth]{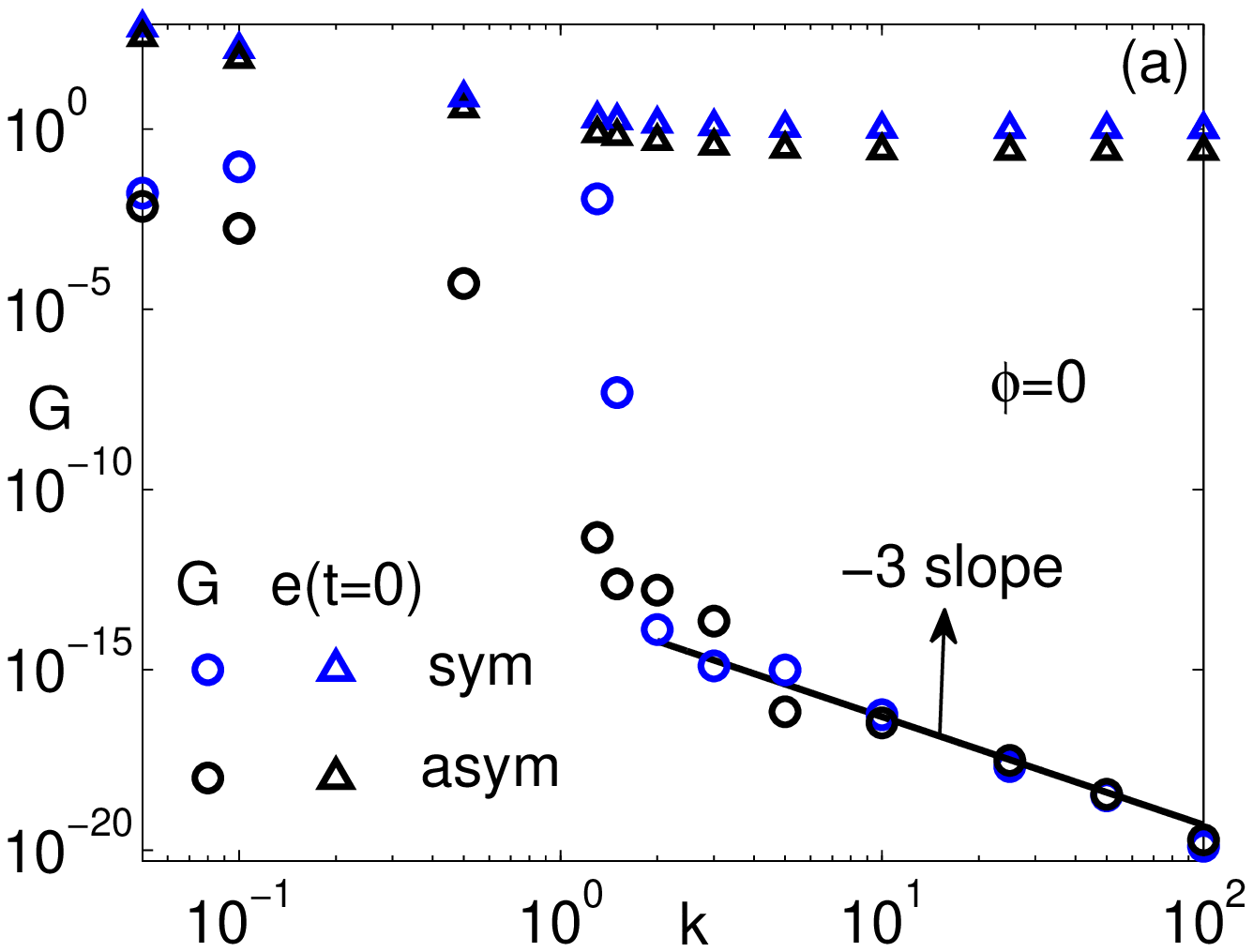}
    \label{E_k_phi_0}
\end{minipage}
\begin{minipage}[]{\columnwidth}
   \includegraphics[width=\columnwidth]{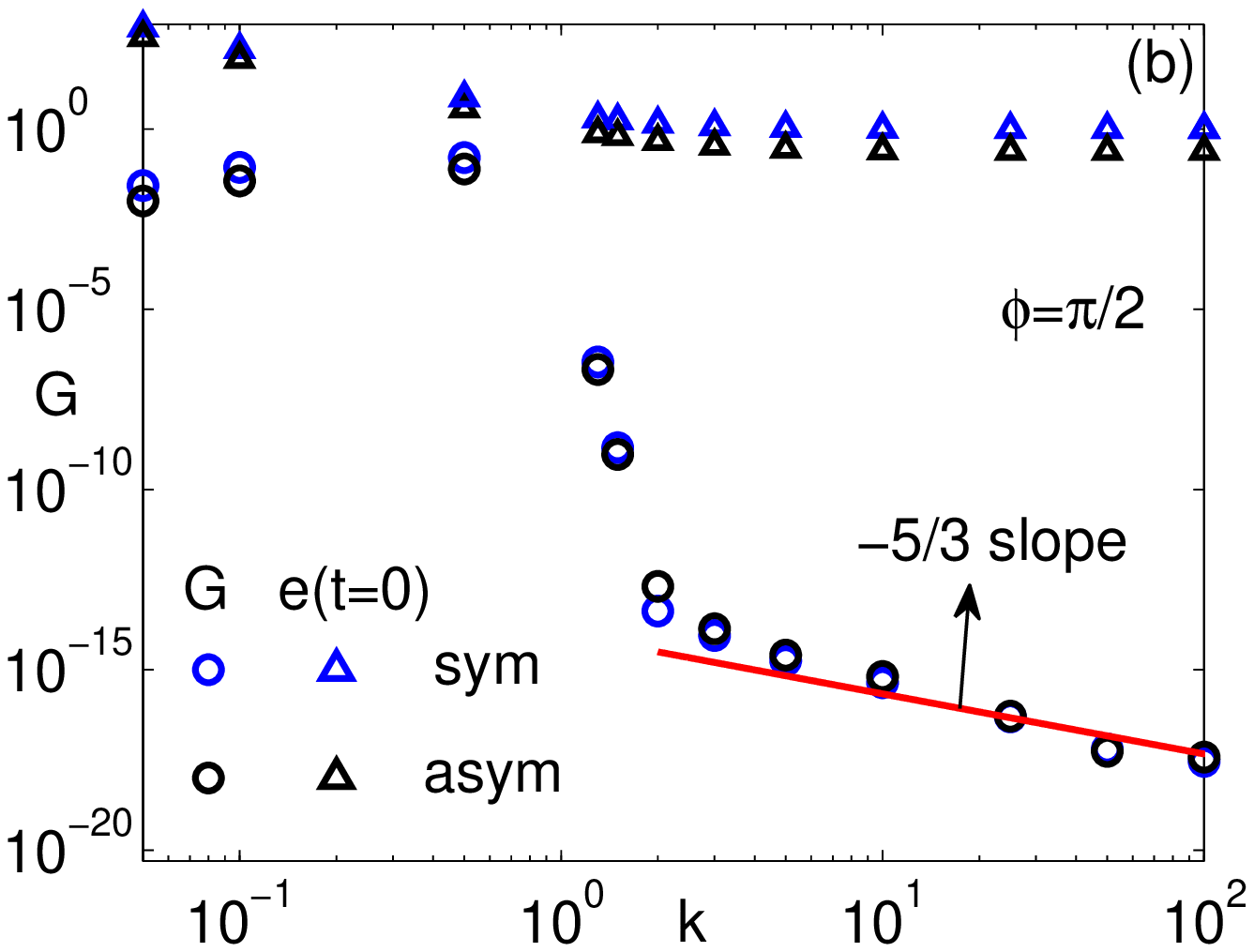}
    \label{E_k_phi_pi_2}
\end{minipage}
\begin{minipage}[]{\columnwidth}
   \includegraphics[width=\columnwidth]{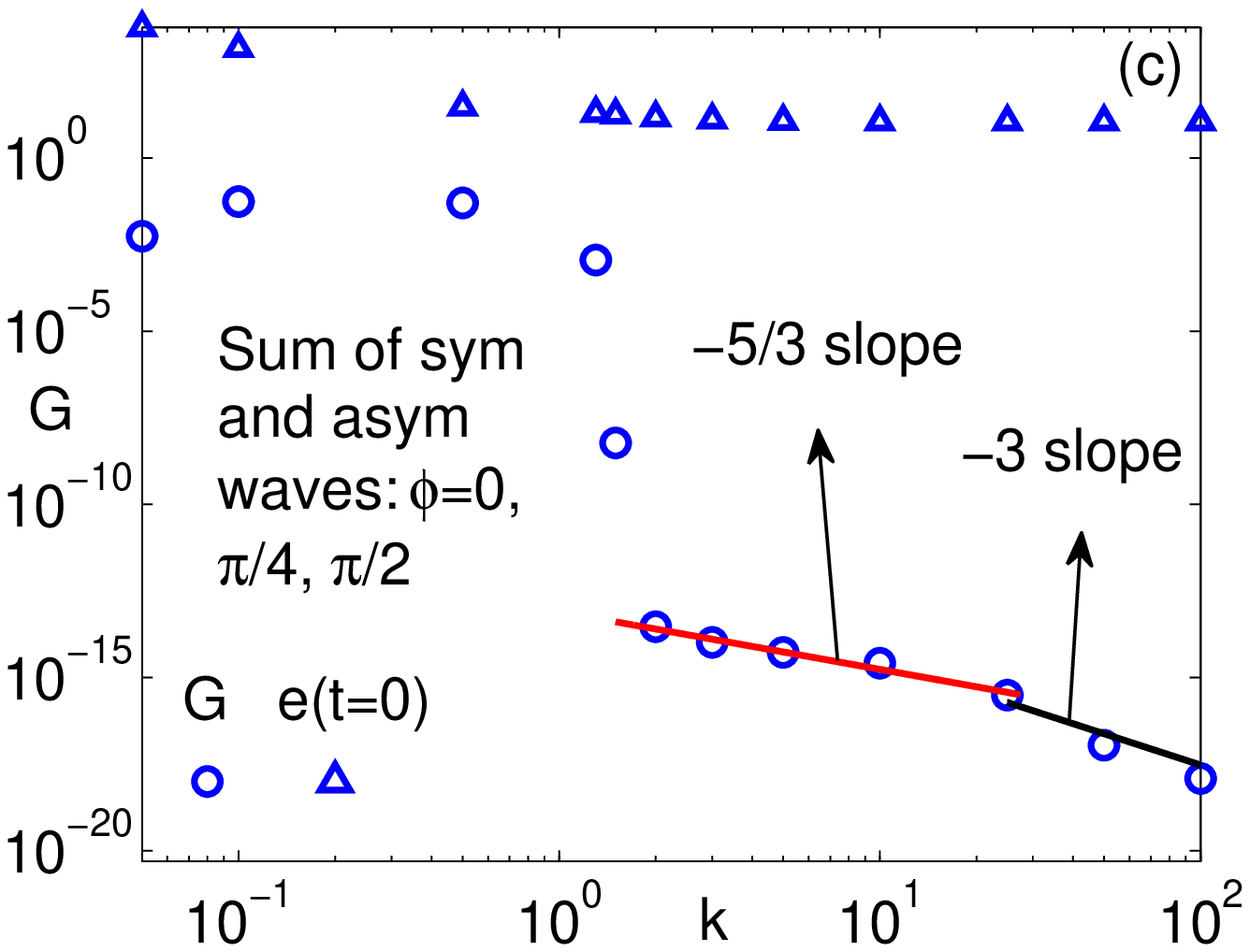}
    \label{E_k}
\end{minipage}
\caption{{\bf Spectrum of the amplification factor of a collection of stable perturbation waves, $Re=40$.} The normalized energy density $G$  at the asymptotic state (circles) and the initial energy density $e(t=0)$ (triangles). The energy of each wave in asymptotic conditions is normalized over the value owned at the initial instant. Blue and black symbols for symmetric and asymmetric initial conditions, respectively. (a) $\phi=0$, (b) $\phi=\pi/2$, and (c) sum of symmetric and asymmetric perturbations with angle of obliquity $\phi=0, \pi/4, \pi/2$. Red and black curves indicate $-5/3$ and $-3$ slopes, respectively.
Note that the intermediate waves are stable and highly damped perturbations, that reach the asymptotic state after a long lapse of time where they progressively lose their kinetic energy. This explain the sharp fall of energy density at the transition between the long and the intermediate wave ranges. }
\label{E_k}
\end{figure}

Computations to evaluate the energy spectrum
are made, at a fixed wavenumber, by integrating the equations forward in time until the temporal growth
rate $r$ asymptotes to a constant value, i.e. when the transient can be considered as extinguished. We consider a range of values for the polar wavenumber $k$ in the interval [0.05, 100], and three angle of obliquity $\phi=0, \pi/4, \pi/2$. For simplicity's sake, the spatial damping rate $\alpha_i$ is here taken equal to $0$, thus no spatial damping is considered in the longitudinal direction. We account for symmetric and asymmetric initial conditions (see red curves in Fig. \ref{perturbation_scheme}) in terms of the transversal velocity $\hat{v}$, while the transversal vorticity $\hat{\omega}_y$ is initially equal to zero.

The normalized energy density $G$ at the asymptotic state (blue and black circles for symmetric and asymmetric initial conditions, respectively) and the initial energy density $e(t=0)$ (blue and black triangles for symmetric and asymmetric initial conditions, respectively) are shown -- as function of the polar wavenumber $k$ -- in parts (a) and (b) of Fig. \ref{E_k} for a longitudinal ($\phi=0$) and a transversal ($\phi=\pi/2$) perturbative wave, respectively. In Fig. \ref{E_k}c we sum, at a fixed wavenumber, asymmetric and symmetric perturbations with angle of obliquity $\phi=0, \pi/4, \pi/2$, and then we calculate the energy density of the resulting perturbation obtained by combining six different waves. We report $-5/3$ and $-3$ slopes with red and black curves, respectively.

In the case of longitudinal waves (two-dimensional base and perturbed flow configurations, see Fig. \ref{E_k}a), the normalized energy density $G$ at asymptotic state has a decay of $-3$ in the inertial range ($k\in[2,100]$) for both symmetric and asymmetric perturbations (see blue and black circles). For smaller wavenumbers, instead, the decay is faster. For purely transversal waves (Fig. \ref{E_k}b), we observe a decay of $-5/3$ in the inertial range ($k\in[2,100]$) for both symmetric and asymmetric perturbations (blue and black circles). Longer waves have a deeper decay, reaching a maximum of energy at about $k=0.5$. If we sum symmetric and asymmetric waves with different angles of obliquity ($\phi=0, \pi/4, \pi/2$), the normalized energy density $G$ (blue circles) of the resulting perturbation at the asymptotic state has a decay of $-3$ for $k\in[25,100]$, while the decay is of the order $-5/3$ for $k\in[2,25]$. For longer waves, the decay is faster and a maximum of energy can be found for $k\in[0.1,0.5]$.

From Eq. (\ref{kinetic_energy}) and recalling that the transversal vorticity $\hat{\omega}_y$ is initially imposed equal to zero, one can observe that the initial energy density has a well-defined decay, given by:
\begin{equation}
e(t=0,k)=\frac{c_1}{k^2} + c_2, \label{initial_energy}
\end{equation}
\noindent where
\begin{equation}
c_1=\int_{-y_d}^{+y_d} \left|\frac{\partial \hat{v}(y,t=0)}{\partial
y}\right|^2 dy,\,\,\,\, c2=\int_{-y_d}^{+y_d} |\hat{v}(y,t=0)|^2 dy.
\end{equation}
This analytical behaviour is confirmed for the three cases we considered in Fig. \ref{E_k}. In parts (a) and (b) blue and black triangles (for symmetric and asymmetric initial conditions, respectively) represent the initial energy density and show a constant value for large wavenumbers ($k\in[2, 100]$), while a decay of about $-2$ is visible for longer waves ($k\in[0.05, 2]$). It should be noted that $e(t=0)$ does not depend on the perturbation obliquity, thus the energy density values at $t=0$ are exactly the same for both $\phi=0$ and $\phi=\pi/2$. In case (c), where six different waves are summed and then the energy density of the resulting perturbation is computed, $e(t=0)$ (blue triangles) still displays a trend in agreement with Eq. (\ref{initial_energy}).

\section{Conclusions}

This is a preliminary study of the behaviour of a collection of transient waves seen throughout their power spectrum. To avoid the complexity introduced by the temporal divergence associated to asymptotically unstable waves, we considered an ensemble of transients of asymptotically stable waves. Then, we have built the energy spectrum by freezing each wave at the moment it reaches a constant (negative) value of the temporal growth rate. When this happens, the waves are dynamically out of their transient and in the asymptotic condition. In the power spectrum, the energy of each wave is normalized over its own initial value. In practice, it is the same as considering the spectral dynamics of a white noise perturbation, a kind of model for the swarm of small perturbations that affects any system in a linear way.

We observe that, whether the waves are aligned with the base sheared flow or not, the energy of the intermediate range of wavenumbers in the spectrum decays with the same exponent ($-5/3$ for 3D oblique waves, $-3$ for 2D aligned waves) that is observed in the spectrum of the velocity fluctuation of fully developed turbulent flows. Where the nonlinear interaction is considered dominant.

At the moment, we can conclude by observing that the spectral power-law scaling of intermediate/inertial waves (with an exponent close to $-5/3$ in 3D and to $-3$ in 2D) is a general dynamical property which encompasses the nonlinear interaction. In other words, it seems to us that the strength of the nonlinearity has been overestimated in the determination of a few turbulence properties.


\end{document}